\documentclass[pre, twocolumn, letterpaper]{revtex4}
\usepackage{graphicx}
\begin{document}

\title{Fluctuation Theorems of Brownian Particles Controlled by a Maxwell's Demon}
\author{Kyung Hyuk Kim}
\email{kkim@u.washington.edu}
\affiliation{Department of Physics, University of Washington, Seattle, 
WA 98195}
\author{Hong Qian}
\email{qian@amath.washington.edu}
\affiliation{Department of Applied Mathematics, University of Washington, 
Seattle, WA 98195}
\date{\today}

\begin{abstract}
We study the stochastic dynamics of Brownian particles in a heat 
bath and subject to an active feedback control by an external, Maxwell's demon-like  agent.  The agent uses the information of the velocity of a particle and reduces its thermal agitation by applying a force.  The entropy of the particle and the heat bath as a whole, thus, reduces.  Entropy pumping [{\em Phys. Rev. Lett.} {\bf 93}, 120602 (2004)] quantifies the entropy reduction.
We discover that the entropy pumping has a dual role of work and 
heat contributing to free energy changes and entropy production of the open-system with the feedback control.  Generalized Jarzynski equality and fluctuation theorems for work functional and entropy production are developed with the presence of the entropy pumping. 
\end{abstract}

\maketitle

\newcommand{\mb}[1]{\mathbf{#1}}
\newcommand{\mbh}[1]{\mathbf{\hat{#1}}}
\newcommand{\Gammah}{\hat{\Gamma}}

\newcommand{\calL}{{\cal{L}}}

Modern nano-technology allows an active control of the position and velocity of nano-devices by feedback systems. The systems detect the velocity and position of the nano-devices and manipulate a velocity- and position-dependent external force applied to the devices. We name such controls as position-dependent feedback control (PFC) and velocity-dependent feedback control (VFC), respectively.    The latter has recently been accomplished in reducing the thermal noise of a cantilever in atomic force microscopy (AFM) \cite{Liang} and dynamic force microscopy \cite{Tamayo}. In \cite{Liang}, the feedback system detects the velocity of the cantilever and reduces its thermal vibration by actively switching the direction and magnitude of a force that controls the cantilever.  A VFC experiment has also been proposed to control a frictional force acting on a small array of particles by limiting the terminal velocity of the array with a terminal attractor \cite{Braiman}.  The physics principle of the VFC is to reduce the entropy of nano-devices and its surrounding environment, as a whole.  The control agent can be considered as a Maxwell's demon, who gathers informations of nano-devices and use them to reduce the entropy.  The entropy reduction has also been widely studied in the context of engineering control theory \cite{Touchette00} and has long
been suggested as a fundamental principle of life \cite{Schrodinger}.

In \cite{Kim03}, as a simple model for the entropy reduction, we studied a Brownian particle in heat bath under a \emph{friction-like} force manipulated by a VFC.  As the system evolves with time, the particle is eventually settled in its stationary state.  In the stationary state, the  average kinetic energy of the particle is lower than the surrounding heat bath  since the fiction-like external force reduces the thermal fluctuations of the particle. Kinetic energy is transferred from heat bath to the particle and is ultimately absorbed by the external control agent.  The  entropy of the heat bath, therefore, decreases in the stationary state, and the particle and the control agent act as a microscopic refrigerator.  To quantify the entropy reduction, a mesoscopic thermodynamic theory has been developed in \cite{Kim03} where we have shown that the entropy of the particle and the heat bath changes due to both  positive entropy production and entropy pumping by the external agent \cite{Kim03}:
\[
\frac{ d (S +  S_H)}{dt}  
	= \frac{dS_p}{dt}+ \frac{dS_{pu}}{dt},
\]
where  $dS$ and $dS_H$ are entropy change in Brownian particles and heat bath, respectively, and $dS_p$ is entropy production  and $dS_{pu}$ is  entropy pumping by a control agent.  	In the case of the friction-like control, $dS_{pu}$ is negative.    In its stationary state, $dS_p+dS_{pu}$ becomes negative.  We note that the entropy production, $dS_p$, is always positive due to the irreversible process of Brownian dynamics and this implies the second law of thermodynamics.   

The entropy reduction mechanism by the  entropy pumping is a unique feature of VFC.  Without the VFC but only with PFC, the entropy pumping term vanishes \cite{Kim03} and the entropy of the particle and the heat bath, as a whole, always increases due to the positive entropy production.  In its stationary state, the entropy of the particle does not change, so  the entropy of the heat bath always increases.
This implies that in the stationary state with PFC alone, heat transfers on average from the system to the heat bath due to the frictional dissipation of work done to the particle by the external agent.  Therefore, the particle and the control agent cannot act as a microscopic refrigerator but as a heater without VFC! 
The Jarzynski equality and fluctuation theorems for entropy production and work functional \cite{Crooks99, Crooks00, Hummer01, Seifert05, Lebo, Kurch98, Jarzynski97, Jarzynski97-pre, Jarzynski00} have been studied only in systems with the PFC.   Thus a novel question arises:  How are they modified with VFC?   The present work answers this question.  Let us formulate this question by focusing on the Jarzynski equality.  The second law of thermodynamics is mathematically an inequality: the external work done on a system in contact with a heat bath, $\langle W \rangle $, is no smaller than the Helmholtz free energy change of the system, $\Delta F$.  This inequality can be
quantified through the newly discovered Jarzynski equality: $\langle e^{-\beta W}\rangle = e^{-\beta \Delta F}$ \cite{Jarzynski97, Jarzynski97-pre, Jarzynski00, Crooks99, Crooks00, Hummer01, Seifert05}, where $\beta \equiv 1/k_BT_H$ with $T_H$ heat bath temperature.  With the presence of VFC, however, the equality needs modification due to entropy pumping. 
Let us consider energy balance: $ \langle W\rangle  = \langle \Delta H \rangle  + \langle Q \rangle $, where $\Delta H$ is the internal mechanical energy change of the system and $Q$ is the heat dissipation from the system to the surrounding heat bath. 
Since the heat bath is in a quasi-static process, $ \langle W \rangle = \langle \Delta H  \rangle + T_H  \Delta S_H   =  \langle \Delta H \rangle + T_H (\Delta S_p + \Delta S_{pu} - \Delta S)$, with $\Delta S_p$ the entropy production,  $\Delta S_{pu}$ the entropy pumping, and $\Delta S$ the entropy change in the system.  Then, $ \langle W \rangle > \langle \Delta H \rangle +T_H( \Delta S_{pu}  -  \Delta S ) = \Delta F + T_H  \Delta S_{pu} $, where $\Delta F \equiv \langle \Delta H \rangle - T_H \Delta S $ is free energy change.  Finally, we get inequality: $
\langle W\rangle - T_H  \Delta S_{pu}  > \Delta F.$
This implies that, with the presence of
VFC, entropy pumping modifies the stochastic work functional in the 
Jarzynski equality.  The 2nd law of thermodynamics can be quantitatively 
described by entropy production fluctuation theorem \cite{Crooks99, Crooks00, Seifert05} and work fluctuation theorem \cite{Lebo, Kurch98}.  These theorems are closely related to Jarzynski equality.  So, they also need to be extended.

We discover that entropy pumping has a dual role of heat and work contributing to free energy change and the fluctuation theorems.  The former role is already mentioned in the previous paragraph.  The latter role can be seen from entropy balance: $dS_p = dS+d(S_H - S_{pu})$. The second law of thermodynamics states that the entropy production $dS_p$ is positive.  Then, in the stationary state, $d(S_H - S_{pu})>0$.  This implies that entropy pumping modifies the heat bath entropy contributing to the fluctuation theorems.

\emph{Langevin Equation --} 
Without losing generality, we consider one-dimension Brownian dynamics described by the following Langevin equation:
\begin{equation}
\frac{dv}{dt} = -\frac{\partial H(x,v;\alpha(t))}{\partial x}-\gamma v + g(v) + \xi,
\label{dvdt}
\end{equation}
with $v$ the velocity to a particle, $\gamma$ the frictional coefficient, $g(v)$ a general VFC, and $\xi$ Gaussian white noise satisfying $\langle \xi(t) \xi(s) \rangle = \delta(t-s)$.  $H(x,v,;\alpha(t))$ is a time-dependent Hamiltonian changing with a parameter $\alpha(t)$ varying with time: $H(x,v;\alpha(t)) = \frac{1}{2}v^2 + U(x;\alpha(t))$. We use unit mass and assume that the Einstein relation $T_H = 1/\gamma$ holds for the heat bath with $T_H$ heat bath temperature.  The corresponding Fokker-Planck equation becomes
$
\frac{\partial P(x,v,t)}{\partial t} = \calL P(x,v,t),
$
where
\begin{equation}
\calL \equiv \partial_v^2- \partial_v[-\{\partial_x H(x,v;\alpha(t))\}-\gamma v + g(v)] - v \partial_x,
\label{calL}
\end{equation}
with $\partial_v \equiv \partial/\partial v$ and $\partial_x \equiv \partial/\partial x$.

\emph{Thermodynamics --}
We first define several terms.  An \emph{internal} system is the Brownian particle together with the surrounding heat bath.  An \emph{external} system is the external agent that manipulates both the control force $g(v)$ and internal potential change due to the change of $\alpha(t)$.

We define stochastic heat $dQ(t)$ \cite{Seki97, Kim03, Seifert05}, stochastic entropy of the Brownian particle $dS(t)$ \cite{Seifert05}, and stochastic entropy pumping $dS_{pu}(t)$ (in the introduction, the same notations were used, but hereafter $S$, $S_p$, and $S_{pu}$ are stochastic.) \cite{Kim03}:
\begin{eqnarray}
dQ(t) &\equiv& -( -\gamma v_t + \xi(t))dx_t \label{dQ}\\
dS(t) &\equiv& -d\ln P(x_t,v_t,t) \label{dS}\\
dS_{pu}(t) &\equiv& \partial_{v_t} g(v_t) dt. \label{dS_pu}
\end{eqnarray}
These are all stochastic quantities since $(x_t,v_t)$ has a stochastic 
trajectory.  The entropy change in heat bath is given due to its 
isothermal quasi-static nature, 
\[
dS_H(t) = \beta dQ(t),
\]
with $\beta = 1/T_H$.
The entropy balance is expressed by 
\begin{equation}
dS+dS_H = dS_p + dS_{pu}.
\label{ent-balance}
\end{equation}
The Eq.(\ref{ent-balance}) can be considered as the definition of  stochastic entropy production, $dS_p$.  Entropy change of the internal system is due to not only entropy production but also entropy pumping. 
Finally, energy balance is expressed as 
\begin{equation}
dH = \partial_t H dt + \partial_x H dx + \partial_v H dv = \partial_t H dt + g dx -dQ,
\label{energy-bal}
\end{equation}
using Eq.(\ref{dvdt}).
So, it is natural to define work done on the particle by external agents, 
\begin{equation}
dW\equiv \partial_t H dt + g dx.
\label{work-balance}
\end{equation}
Note that all the above stochastic thermodynamic quantities are defined with Stratonovich prescription, which is known to be physically meaningful \cite{Kim05}.

\emph{Jarzynski equality and fluctuation theorems --}
The Jarzynski equality \cite{Jarzynski97, Jarzynski97-pre, Jarzynski00, Hummer01} and the equality due to entropy production fluctuation theorem \cite{Crooks99, Crooks00, Seifert05}  can be derived by examining the temporal behavior of the  following quantity:
\[
\lefteqn{f(x,v,t) = \Big\langle \delta(x-x_t) \delta(v-v_t)} 
\hspace{4in}
\]
\begin{equation}
	\times \exp\left[\int_{s=0}^{s=t} -dS_H(s) + dS_{pu}(s) 
		+ d\ln w(x_s,v_s,s)\right]\Big\rangle, 
\label{f}
\end{equation}
where $w(x,v,s)$ is an arbitrary weight function.
Note that 
$\langle \cdots \rangle$ is a path integral averaging over initial distribution, $P(x,v,0)$: 
\begin{eqnarray*}
\lefteqn{\langle \cdots \rangle  \equiv \int \lim_{N \rightarrow \infty}\prod_{i=0}^{N} dx_idv_i (\cdots) P(x_N,v_N|x_{N-1},v_{N-1})} \\
	&&\times P(x_{N-1},v_{N-1}|x_{N-2},v_{N-2}) \cdots P(x_1,v_1|x_0,v_0) P(x_0,v_0,0),
\end{eqnarray*}
where $P(x,v|x^\prime, v^\prime)$ is transition probability to find a Brownian particle at $x$, $v$ after a time interval $\epsilon \equiv t/N$ given $x^\prime$, $v^\prime$ as initial starting point.
$-dS_H + dS_{pu}$ can be expressed as
$\beta dH - \beta \partial_t H dt - \beta g dx + \partial_v g dt$ using Eq.(\ref{dvdt}), (\ref{dQ}), and (\ref{dS_pu}).
Eq.(\ref{f}) becomes
\begin{eqnarray*}
f(x,v,t) &=& w(x,v,t) e^{\beta H(x,v;\alpha(t))} f_0(x,v,t),
\end{eqnarray*}
where
\begin{eqnarray*}
\lefteqn{f_0(x,v,t) \equiv\Big\langle\frac{\delta(x-x_t)\delta(v-v_t)}{w(x_0,v_0,0)e^{\beta H(x_0,v_0;\alpha(0))} } }\\
	&&\times \exp\left[\int_{s=0}^{s=t} -\beta(\partial_s H(s)ds + g(v_s)dx_s)+ \partial_{v_s} g(v_s) ds\right]\Big\rangle.
\end{eqnarray*}
Note that $f_0(x,v,0) = \frac{P(x,v,0)}{w(x,v,0)}e^{-\beta(H(x,v;\alpha(0)))}$.
The time derivative of $f_0(x,v,t)$ is expressed as
\[
\partial_t f_0(x,v,t) = \calL f_0(x,v,t) + f_0(x,v,t) [- \beta \partial_t H - \beta g v + \partial_v g].
\]
Its solution becomes $f_0(x,v,t) = e^{-\beta H(x,v;\alpha(t))}$ by requiring $w(x,v,0) = P(x,v,0)$.  Therefore, $f(x,v,t) = w(x,v,t)$.  By integrating both the sides of Eq.(\ref{f}) over $x$ and $v$, we get the following general equalities \cite{Seifert05},
\begin{equation}
\Big\langle \frac{w(x_t,v_t,t)}{P(x_0,v_0,0)} \exp[-\Delta S_H(t) 
		+ \Delta S_{pu}(t)] \Big\rangle =1,
\label{f-theorem}
\end{equation}
where  $w(x,v,t)$ is an arbitrary weight function with $w(x,v,0) = P(x,v,0)$, and $P(x,v,0)$ is an arbitrary initial probability distribution.

With $w(x,v,t) = \exp[-\beta H(x,v;\alpha(t))]/Z_e(t)$, where $Z_e(t) \equiv \int dxdv \exp[-\beta H(x,v;\alpha(t))]$, Eq.(\ref{f-theorem}) becomes an extended form of Jarzynski-equality:
\begin{equation}
\big\langle e^{-\beta W(t)+\Delta S_{pu}(t)}\big\rangle = e^{-\beta \Delta F(t)},
\label{jar-equal}
\end{equation}
where, using Eq.(\ref{work-balance}), 
\[
W(t) \equiv \int_0^t ds \Big[\frac{\partial H(x_s,v_s;\alpha(s))}{\partial s} + g(v_s)v_s\Big]
\]
is the work done on the particle by external control agents, and  $
\Delta F(t) \equiv -\ln \frac{Z_e(t)}{Z_e(0)}$,
is the free energy difference of two equilibrium state parameterized by $\alpha(0)$ and $\alpha(t)$, respectively.  
Note that the final probability distribution does \emph{not} have to be in equilibrium states parameterized by $\alpha(t)$, while the initial one does by parameter $\alpha(0)$.

With $w(x,v,t) = P(x,v,t)$, Eq.(\ref{f-theorem}) becomes an extended form of an equality related to the entropy production fluctuation theorem \cite{Seifert05},
\begin{equation}
\langle \exp[-\Delta S_H(t) -\Delta S(t) + \Delta S_{pu}(t)] \rangle= \langle \exp[-\Delta S_p(t)] \rangle =1.
\label{equal-epft}
\end{equation}
Eq.(\ref{equal-epft}) shows that  average entropy production $\langle \Delta S_p(t) \rangle $ becomes positive over the finite time interval with  or without VFC for arbitrary initial distribution, $P(x,v,0)$.  Eq.(\ref{equal-epft}) also implies that entropy production fluctuation theorem holds under VFC with proper definition of $S_p$, Eq.(\ref{ent-balance}).

With $w(x,v,t) = w(x,v,0)$, we obtain a novel equality,
\begin{equation}
\langle \exp[-\Delta S_H(t) + \Delta S_{pu}(t)] \rangle= 1,
\label{equal-hdft}
\end{equation}
over flat initial distribution ($P(x,v,0)=$ constant). Without VFC, Eq.(\ref{equal-hdft}) becomes $\langle \exp[- \Delta S_H (t)] \rangle =1$ indicating that, over the \emph{finite} time interval,  the average heat dissipation $\langle \Delta S_H \rangle $ is guaranteed to be positive without VFC for flat initial distribution only. 

We note that, for different initial probability distributions, one can get various equalities while an equality related to entropy production fluctuation theorem is independent of initial probability distributions.

\emph{Entropy production fluctuation theorem --}
We now obtain an extended form of entropy production fluctuation theorem:
\begin{equation}
\frac{P(\Delta S_p(t)=a)}{P(\Delta S_p(t)=-a)}=\exp[a],
\label{epft}
\end{equation}
using the following path integral relation \cite{Crooks99},
\begin{equation}
\frac{P(\{x_s,v_s\};\alpha(s))}{P(\{x_{t-s},-v_{t-s}\};\alpha(t-s))} = \exp[\Delta S_H(t) - \Delta S_{pu}(t)],
\label{strong-gcft}
\end{equation}
where $P(\{x_s,v_s\};\alpha(s))$ is the probability to find a path, $\{x_s, v_s \}$, with $0\leq s \leq t$, starting from $x_0$, $v_0$ and ending at $x_t$, $v_t$ and $P( \{x_{t-s},-v_{t-s}\};\alpha(t-s))$ is the probability to find a path traced backward.  The derivation of Eq.(\ref{strong-gcft}) is based on the following conditional probability ratio:
\begin{eqnarray*}
\frac{P( x, v | x^\prime, v^\prime )}{P(x^\prime, -v^\prime | x,-v)} &=& \frac{\langle x, v | e^{\epsilon \calL}|x^\prime, v^\prime \rangle}{\langle x^\prime, -v^\prime | e^{\epsilon \calL}|x, -v \rangle},
\end{eqnarray*}
where ${\calL}$ is Fokker-Planck operator defined as Eq.(\ref{calL}).
 To make the transition probability into path integral form, we express $\calL$ into a Weyl-ordered form:
\begin{eqnarray*}
\lefteqn{\calL_w(x,v,\hat{p}_x,\hat{p}_v) = -\hat{p}_v^2 - i v \hat{p}_x - \frac{1}{2}i \hat{p}_v[F(x,v) - \gamma v ]} \\
	&&- \frac{1}{2}[F(x,v) - \gamma v ]i\hat{p}_v - \frac{1}{2}\Big( \partial_v[F(x,v) - \gamma v ] \Big),
\end{eqnarray*}
where $F(x,v;\alpha(t)) \equiv -\partial_x H(x,v;\alpha(t)) + g(v)$, $\hat{p}_x \equiv -i \partial_x$ and $\hat{p}_v \equiv -i \partial_v$.
Then, as $\epsilon \rightarrow 0$, 
\begin{eqnarray*}
\lefteqn{P( x, v | x^\prime, v^\prime ) =\int dp_x dp_v \exp[ \epsilon \calL_w(\bar{x},\bar{v},p_x, p_v)}\\
	&&+ip_x(x-x^\prime)+ ip_v(v-v^\prime)] \\
	&=& \frac{\delta(x-x^\prime -  \epsilon \bar{v})}{\sqrt{4 \pi \epsilon}} \\
		&&\times \exp\Big[  - \epsilon \Big( \frac{\bar{F}-\gamma \bar{v}}{2} - \frac{v- v^\prime}{2\epsilon}\Big)^2 - \frac{\epsilon}{2}[\partial_{\bar{v}}(\bar{F}-\gamma \bar{v})]  \Big],
\end{eqnarray*}
and
\begin{eqnarray*}
\lefteqn{P(x^\prime, -v^\prime|x,-v) = \frac{\delta(x-x^\prime -  \epsilon \bar{v})}{\sqrt{4 \pi \epsilon}}} \\
	&& \times \exp\Big[  - \epsilon \Big( \frac{\bar{F}+\gamma \bar{v}}{2} - \frac{v- v^\prime}{2\epsilon}\Big)^2 + \frac{\epsilon}{2}[\partial_{\bar{v}}(\bar{F}+\gamma \bar{v})]  \Big],
\end{eqnarray*}
where $\bar{x}\equiv (x+x^\prime)/2$, $\bar{v} \equiv (v+v^\prime)/2$, and $\bar{F} \equiv F(\bar{x}, \bar{v})$. Therefore, 
\begin{eqnarray*}
\frac{P(x,v|x^\prime, v^\prime)}{P(x^\prime, -v^\prime|x,-v)}&=&\exp \Big[  \epsilon  \Big( \bar{F} - \frac{v-v^\prime}{\epsilon} \Big) \gamma \bar{v}    -\epsilon \partial_{\bar{v}}\bar{F}  \Big],\\
	&=& \exp[dS_H - dS_{pu}].
\end{eqnarray*}

\emph{Work Fluctuation Theorem --}
When the Hamiltonian is time-independent, the work fluctuation theorem has been obtained \cite{Kurch98, Lebo}.  Like the entropy production fluctuation theorem, the work fluctuation theorem is extended as follows.  From energy balance, $\Delta S_{p}(t)= -\beta \Delta H(t) + \beta W(t) +\Delta S(t) - \Delta S_{pu}(t)]$,
where $W(t) = \int_0^t dx_s g(v_s)$ in the time-independent Hamiltonian case. 
$|\Delta S(t)|$ does not increase with sufficiently large time $t$  on average since $\Delta S(t) = -\ln P_{ss}(x_t,v_t) + \ln P(x_0,v_0,0)$ with $P_{ss}$ a stationary distribution as $t \rightarrow \infty$, while $|W(t)|$ and $ |\Delta S_{pu}|$ increase on average.  Therefore, $\Delta S_p(t)  \rightarrow \beta W(t) - \Delta S_{pu}(t)$ for $t \rightarrow \infty$ and extended work fluctuation theorem holds for $t \rightarrow \infty$:  
\begin{equation}
\frac{P(\beta W(t)-\Delta S_{pu}(t)=a)}{P(\beta W(t)-\Delta S_{pu}(t)=-a)} = \exp[a].
\label{workft}
\end{equation}
The corresponding equality is derived,
\begin{equation}
\lim_{t \rightarrow \infty}\langle \exp[ - \beta W(t) + \Delta S_{pu}(t)]\rangle =1.
\label{equal-workft}
\end{equation}
From Eq.(\ref{equal-workft}), we find that $\langle W(t) - T_H\Delta S_{pu}(t) \rangle >0$ as $t \rightarrow \infty$ with a time-independent Hamiltonian.

\emph{Entropy Pumping --}
In \cite{Kim03}, we have found that entropy pumping is related to momentum phase space contraction due to $g(v)$ ($dS_{pu}/dt = \partial_v g(v)$).  The extended fluctuation theorems derived above show a novel role of entropy pumping: \emph{a dual role of work and heat}.  The work functional $W$ in the Jarzynski equality and the work fluctuation theorem is modified into $W-T_H \Delta S_{pu}$  as in Eq.(\ref{jar-equal}) and (\ref{workft}),  and heat dissipation $Q$, i.e., entropy of heat bath $S_H$ is modified  into $S_H - S_{pu}$ as in Eq.(\ref{f-theorem}), (\ref{equal-epft}), and (\ref{equal-hdft}).   The reason of the duality is easy to understand from energy balance Eq.(\ref{energy-bal}): $\Delta H = W - T_H \Delta S_H$.  When $S_H$ is modified, $W$ also needs to be modified with $\Delta H$ unchanged.  

For definiteness, let us consider and compare three examples with $U(x; \alpha(t)) = 0$: (1) without VFC but only with PFC in two and higher dimensional systems ($g(v)$ in Eq.(\ref{dvdt}) is replaced to $\vec{g}(\vec{x})$.), (2) with friction-like VFC, $g(v) = -c v$ with $c >0$, and (3) with non-friction-like VFC with $c <0$.
Let the system in a stationary state. In the case (1), $\Delta S_{pu}=0$ and the average work by control force (C) onto Brownian particles (BP), $\langle W_{C \rightarrow BP}\rangle $, becomes positive from Eq.(\ref{equal-workft}), so heat dissipation from BP to heat bath (HB), $\langle Q_{BP \rightarrow HB}\rangle $ becomes positive.   In the case (2), $\langle W_{C \rightarrow BP} \rangle < 0$ and $\langle Q_{BP \rightarrow HB}\rangle < 0$.  However, the modified work and heat have opposite signs: $\langle W_{C \rightarrow BP}\rangle -T_H \langle \Delta S_{pu}\rangle > 0$ and $\langle Q_{BP \rightarrow HB}\rangle -T_H \langle \Delta S_{pu}\rangle > 0$, where $\langle \Delta S_{pu} \rangle = -c \Delta t < 0$  \cite{Kim03}.  With the modified work and heat, the case (2) becomes the case (1)! 
In the case (3), $\langle W_{C \rightarrow BP}\rangle >0$ and $\langle Q_{BP \rightarrow HB}\rangle > 0$. With entropy pumping modification, $\langle W_{C \rightarrow BP}\rangle -T_H \langle \Delta S_{pu}\rangle >0$ and $\langle Q_{BP \rightarrow HB}\rangle -T_H \langle \Delta S_{pu}\rangle > 0$, where $\langle \Delta S_{pu} \rangle = -c \Delta t > 0$.

\emph{Conclusion Remarks --}
Nano-scale mesoscopic systems with VFC are 
significantly different from the widely studied overdamped stochastic 
systems with PFC \cite{QianJPCM05}. The key difference is that the 
former involves an active entropy reduction mechanism like a Maxwell's Demon.  We have recently developed a rigorous mesoscopic 
thermodynamic theory for such systems with VFC \cite{Kim03}: from the standpoint
of the first law of thermodynamics, a friction-like VFC makes the heat flow into Brownian particles from the heat bath and then be taken out to the external control agent in the form of  mechanical energy.  However, from the standpoint of the second law, not all the heat can be converted into mechanical energy.
Otherwise,  we would have a perpetual motion machine of the 
second kind. This is the paradoxical part of a Maxwell's Demon.  
This paradox is resolved by considering the Brownian particles and demon-like external agent as one whole system.  
Since the Brownian particles cannot be treated as one whole system, a question, `what constitutes the
\emph{stochastic work} contributing to the change in free energy of the Brownian particles?', arises.  In this 
Letter, we have shown that the stochastic work and entropy production appearing in Jarzynski equality and fluctuation theorems should include entropy pumping contribution due to a dual role of entropy pumping as work and heat. 
With this modification, we are able to extend all the previous results
to systems with VFC. 

\emph{Acknowledgment --}
We thank M. den Nijs and S. Yoon for useful discussions and comments.  	 
This research is supported by NSF under grant DMR-0341341.
\bibliography{ft}
\end{document}